\documentclass[12pt,a4paper]{article}

\title{{\large\textbf{Upper bounds on High-Scale Supersymmetry in the singlet-extended MSSM}}}
\author{Lucila Z\'arate}
\date{}

\usepackage{amsmath, amssymb, amsfonts, mathrsfs, graphicx, fancyhdr}
\usepackage[square,sort,comma,numbers]{natbib}
\usepackage{setspace}
\def\be{\begin{equation}}
\def\ee{\end{equation}}
\def\ba{\begin{align}}
\def\ea{\end{align}} 

\begin{document}

\begin{titlepage}


\begin{center}
{
\large\textbf{The Higgs mass and the scale of SUSY breaking \\in the NMSSM}}

\vspace{1cm}

\textbf{
Lucila Zarate
}
\\[5mm]
{\small 
II. Institut f\"ur Theoretische Physik der Universit\"at Hamburg, Luruper Chaussee 149,
D-22761 Hamburg, Germany
}
\end{center}

\vspace{1cm}

\begin{abstract}

In this letter we study the Higgs mass in the NMSSM 
with supersymmetry breaking at high scales $M_{SS}$. 
With the Standard Model as the effective low energy theory, 
the computation of the Higgs mass 
relies on the matching condition of the quartic coupling $\lambda$ at $M_{SS}$.
In the MSSM, the latter is fixed to a semi-positive 
value and, thus, sets an upper bound on the SUSY-breaking scale near $M_{SS}\simeq 10^{10}\text{GeV}$.
In the NMSSM, $\lambda(M_{SS})$ receives an additional contribution induced by the singlet which allows for negative values of $\lambda(M_{SS})$. 
In turn, 
for the measured value of the Higgs mass we find that $M_{SS}$ can take any value up to the GUT scale.
Furthermore, the choice of universal soft terms favors 
SUSY-breaking scales close to the GUT scale.


\end{abstract}
\today

\end{titlepage}

\section{Introduction}

The discovery of a Higgs boson of $126\text{GeV}$ \cite{Chatrchyan:2013mxa, Khachatryan:2014ira, Aad:2014aba} is consistent with low energy supersymmetry together with a small degree of fine-tuning. However the absence of supersymmetric particles in the $\text{TeV}$ range questions the paradigm of naturalness as a guiding principle. 
If one abandons the idea of supersymmetry as a solution to the hierarchy problem,  reasons remain to study realizations of supersymmetry. In particular, supersymmetry is present in string theory, which is so far the best candidate for a UV-complete theory.  
Generically, supersymmetric theories in four dimensions obtained from string compactifications
favor supersymmetry breaking at high scales, e.g.~the GUT scale \cite{Douglas:2006es}.\footnote{Examples with low energy SUSY were constructed by special choices of the compactification data (e.g.~fluxes).} 
Furthermore, present data confirmed that the SM Higgs potential becomes metastable at large energy scales\cite{Buttazzo:2013uya} and supersymmetric embeddings can stabilize the EW vacuum.\footnote{See \cite{Ellis:2009tp,EliasMiro:2011aa,Degrassi:2012ry} for earlier results on the stability of the EW potential.} 

 
Let us consider the situation in which supersymmetry is broken at a scale $M_{SS}$ such that $M_{EW}\ll M_{SS}$. Under no special assumptions on the corresponding pattern of soft terms, supersymmetric particles receive masses of $\mathcal{O}(M_{SS})$ and thus decouple from the low energy theory. The requirement of obtaining a light Higgs in the spectrum leads to a fine-tuning condition 
at the matching scale $M_{SS}$ and leaves the Standard Model as an effective description at lower energies. However, the corresponding quartic coupling $\lambda(\mu)$ is completely fixed by the matching conditions to the supersymmetric theory at the scale $M_{SS}$.\footnote{Here $\mu$ denotes the energy scale.} In turn, the latter implies that the Higgs mass is determined by the supersymmetric theory,
and thus, the measured value of the Higgs mass constrains
the values of the parameters of the supersymmetric theory and $M_{SS}$. 
This framework is called High-Scale Supersymmetry. 
A detailed study of the Higgs mass in the High-Scale Supersymmetry within the MSSM was done in \cite{Hall:2009nd, Giudice:2011cg, Cabrera:2011bi, Arbey:2011ab, Draper:2013oza, Bagnaschi:2014rsa}. 
In this case the prediction of $\lambda(M_{SS})$ is strictly positive (or zero), thus, the scale at which the quartic coupling in the SM vanishes, within uncertainties, sets an upper bound on $M_{SS}$ of $M_{SS}\simeq 10^{10}\text{GeV}$. Hence, a GUT scale SUSY breaking in this setup is ruled out.


In this letter we study High-Scale Supersymmetry within the singlet extension of the MSSM, i.e.~Next-to-Minimal Supersymmetric Standard Model (NMSSM), for a comprehensive review see \cite{Ellwanger:2009dp}.  
In the NMSSM the couplings of the extra singlet to the Higgs sector induce tree level corrections to the matching of $\lambda$ at the $M_{SS}$ scale. 
Two pieces contribute to $\lambda(M_{SS})$ that are model independent and model dependent respectively. The former is strictly positive (or zero), 
and it can significantly modify the parameter values consistent with the Higgs mass w.r.t the MSSM case.  
On the other hand, the latter contribution allows to set $\lambda(M_{SS})$ to a negative value, and thus, can extend $M_{SS}$ up to the GUT or Planck scales.\footnote{This possibility was anticipated in \cite{Giudice:2011cg} and proposed also in \cite{Hebecker:2013lha}.}
In this work we calculate the prediction for the Higgs mass in this framework and derive the respective bounds on $M_{SS}$ and the NMSSM parameters.  
 
 This letter is organized as follows in section \ref{settingNMSSM} we write down the matching conditions that relate the singlet-extended MSSM framework with the effective SM and discuss the fine-tuning condition required to induce a light Higgs. In section \ref{calcHiggsmass} we explain how we calculate the Higgs mass and in section \ref{HiggsvsMss} we study the dependence of the Higgs mass as a function of the SUSY-breaking scale and NMSSM parameters. In section \ref{UVdependence}
we study the constraints on the Higgs mass for special choices of soft parameters at the GUT scale. Finally, in section \ref{conclusion} we present the conclusions.
 
\section{Fine-tuning a light Higgs in SUSY and matching to the Standard model}\label{settingNMSSM}

Let us consider a scenario in which above the (not necessarily low) supersymmetry breaking scale ($M_{SS}$) the theory is described by a singlet-extended MSSM with the following superpotential
\be 
W_{\text{NMSSM}}= (\mu_h+y_s S)H_u H_d+\tfrac{1}{2}\mu_s S^2 +
y_u Q
H_uU_R+y_dQH_dD_R+y_eLH_dE_R \,,
\label{Wnmssm}\ee
where $S$ is the NMSSM singlet and $H_u,H_d$ are the MSSM Higgs
multiplets. $y_s$ is a dimensionless Yukawa coupling
and $\mu_h,\mu_s$ are the supersymmetric Higgs and singlet mass terms respectively.
$y_u,y_d,y_e$ are  the Yukawa couplings of the MSSM which should be understood as matrices in the family space, $Q$ are the quark doublets, $U_R$ and $D_R$ are the quark singlets, $L$ are the lepton doublets and $E_R$ are the lepton singlets.
After supersymmetry breaking the scalar potential of the Higgs sector develops soft terms which read
\be
\begin{aligned} 
V_{\text{soft}}(H_u,H_d,S)=& m_{h_u}^2 \vert H_u\vert^2+ m_{h_d}^2 \vert H_d\vert^2+m_3^2\,(H_u H_d+h.c.)\\
&+m_{h_s}^2\vert S\vert^2+(y_s A_{\lambda}S\,H_u H_d+\tfrac{1}{2}b_s^2 S^2+h.c.)\,.
\end{aligned}
\ee
where $m_j$ are the soft scalar masses, $A_{y}$ the $A$-term and $m_3^2,b_s^2$ are the 
$b$-term and bilinear soft term of the singlet respectively.\footnote{It is worth pointing out that, unless they are forbidden by a symmetry, effective soft cubic and linear terms for the singlet can be generated radiatively. As long as $y_s$ is small, the latter are suppressed and can be neglected. Throughout this letter we will not consider them in the calculations.}

The explicit computation of the scalar CP-even Higgs potential shows that the condition, at the scale $M_{SS}$,\footnote{Here we used that the off-diagonal terms in the mass matrix (of the three CP-even scalars) that mix the MSSM Higgses with the singlet can be neglected in the diagonalization. 
}
\be \hat{m}_3^4\simeq (m_{h_u}^2 + \mu^2)(m_{h_d}^2 + \mu^2)\,\label{ft}\ee
with $\mu=\mu_h+y_s \langle S\rangle$ and $\hat{m}_3^2=m_3^2+y_s (A_\lambda+\mu_s)\langle S\rangle$, generates a massless Higgs field given by the combination
\be H_{SM}=\sin \beta H_u-\cos \beta \epsilon H_d^{\ast} \ee
where $\epsilon$ is the antisymmetric tensor and the angle $\beta$ is determined by
\be 
\tan^2\beta= \frac{\vert m_{h_d}^2 + \mu^2\vert}{\vert m_{h_u}^2+\mu^2\vert}
\label{tanbeta}
\ee
with $m_{h_u},m_{h_d}\text{ and }\mu$ evaluated at the scale $M_{SS}$.
It can be easily seen that the choice of superpotential in equation \eqref{Wnmssm} yields $\langle S \rangle\ll m_{EW}$ and thus the singlet contribution to $\mu$ is suppressed and can be neglected.
Moreover, this implies effective quadratic terms are subleading and hence the fine-tuning conditions in \eqref{ft} and \eqref{tanbeta} are identical to the MSSM case.

Assuming the generic situation that the susy particles get masses of $\mathcal O(M_{SS})$ they can be integrated out, leaving an effective Standard Model description at energies below the cutoff scale $M_{SS}$.
Furthermore,
the explicit computation of the effective Lagrangian
 provides the boundary condition for the quartic coupling in the Standard Model potential given in \eqref{V_SM}. These are the so called matching conditions and at tree level they read 
\be 
\lambda^{\text{tree}}(M_{SS})=\tfrac{1}{4}\left(g_2^2+\tfrac{3}{5}g_1^2\right)\cos^2 2\beta + \tfrac{1}{2}y_s^2 (1-\delta) \sin^2 2\beta 
\label{lam}
\ee
where 
\be \delta=\frac{\left(2\mu_h/\sin 2\beta - A_{\lambda}-\mu_s\right)^2}{m_s^2+b_s^2+\mu_s^2}\,.
\label{delta}
\ee
The first term in \eqref{lam} is the well known D-term contribution in the MSSM while the second term appears only in the singlet-extension. The latter is generated by two effects, an F-term contribution generated by the Yukawa interaction that couples the singlet to the Higgs, and an extra contribution (proportional to $\delta$) originated from integrating out the singlet. 
Interestingly, the denominator in \eqref{delta} corresponds to the mass of the (CP-even) scalar singlet and, thus, it is positive. In turn, $\delta$ can only take positive values and the correction to $\lambda(M_{SS})$ is always negative.

The matching given in \eqref{lam} receives higher order threshold corrections  ($\delta\lambda^{th}$) that for the MSSM were originally computed at one loop level in \cite{Giudice:2011cg} 
(and recently reviewed in \cite{Bagnaschi:2014rsa} with leading two loop effect). We follow \cite{Bagnaschi:2014rsa} and parametrize the corrections as follows\footnote{In our setup these are not complete, the threshold corrections coming from integrating out the two scalar singlets and the singlino are not included.}
\be 
\delta\lambda^{th}(M_{SS})=\Delta\lambda^{1l,\text{reg}}+\Delta\lambda^{1l,\phi}+\Delta\lambda^{1l,\chi^{1,2}}\,.
\label{lam_threshold}\ee
These originate from the change of renormalization schemes that relate the gauge couplings in the $\overline{\text{DR}}$ scheme to the $\overline{\text{MS}}$ scheme ($\Delta\lambda^{1l,\text{reg}}$) and from integrating out the heavy scalars ($\Delta\lambda^{1l\phi}$) and fermionic superpartners ($\Delta\lambda^{1l\chi^{1,2}}$). The effect induced from stop mixing is also included, with the stop mixing parameter defined as
\be \tilde{X}_t=(A_t-\mu_h \cot\beta)^2/(m_Qm_U)\,.
\label{stopmixing}
\ee  

Before finishing this section it is worth noticing that, as long as one does not 
assume a special pattern of soft terms, the free parameters that determine $\lambda(M_{SS})$ in \eqref{lam} are
\be 
y_s \,,\delta\,,\tan\beta\,\text{and}\ M_{SS} \,.
\label{freepar}\ee
However, 
if the soft terms are specified, the corresponding soft parameters at $M_{SS}$ completely determine $\tan\beta$ and $\delta$ via \eqref{tanbeta} and \eqref{delta} respectively.



\section{Calculation of the Higgs mass}\label{calcHiggsmass}

The Standard Model Higgs potential reads
\be 
V_{SM}=\tfrac{1}{2}\lambda(\vert H\vert ^2-v^2)^2\,
\label{V_SM}\ee
where $v=\sqrt{2}\cdot174.1\text{GeV}$ is the the Higgs vacuum expectation value.\footnote{In the literature another convention for $v$ is often used, without the squared root of 2. In that case, $M_h^2=2\lambda v^2$, the result is of course independent of this definition.} From \eqref{V_SM} one immediately learns that value the of $\lambda$ determines the Higgs mass $M_h$ at loop level via 
\be 
M_h^2 = v^2(\lambda+\delta_{\lambda})\,.
\label{M_h}
\ee
It is worth recalling that the top Yukawa coupling $y_t$ is fixed at loop level by the mass of the top quark through
\be 
y_t=\frac{m_t}{v}\sqrt{2}(1+\delta_t)
\label{m_t}
\ee
$\delta_\lambda$ and $\delta_t$ parametrize the threshold corrections at the renormalization scale ($m_t$) that were originally computed in \cite{Sirlin:1985ux} and \cite{Chetyrkin:1999qi} respectively. The state-of-the-art computations are given at the two loop level (and dominant three loop correction for $y_t$) in \cite{Buttazzo:2013uya}.

\subsubsection{Matching at $M_{SS}$}\label{matchingMSS}

We perform the numerical calculations using  a modified version of SPheno-3.3.6 \cite{Porod:2003um, Porod:2011nf} created by SARAH-4.5.8 \cite{Staub:2008uz,  Staub:2010jh, Staub:2013tta}. 
Given $\lambda(M_{SS})$ the Renormalization Group Equations (RGEs) are calculated at two-loop-level to yield the couplings at the weak scale. All couplings are renormalized at one loop at $m_t$ in the $\overline{\text{MS}}$ scheme and the corresponding Higgs mass is thus calculated at one loop level. For the top Yukawa coupling we include the  two loop and dominant three loop QCD correction given in eq.57 in \cite{Buttazzo:2013uya}.
For completeness we provide the values of the SM parameters used in the calculations
\be
\begin{aligned}
&m_t=173.34\pm0.76\, \text{GeV}\,\quad \alpha_s=0.1184\\
&M_z=91.18\,\text{GeV}\,,\quad G_F=1.16637\cdot10^{-5}\ .
\end{aligned}
\ee

A theoretical uncertainty on the Higgs mass of $3\,\text{GeV}$ is generically applied to supersymmetric models. This was computed within the MSSM in \cite{Degrassi:2002fi}, assuming low energy values of the SUSY-breaking scale. The computation was recently reviewed in \cite{Vega:2015fna}, for arbitrary (large) values of the SUSY-breaking scale and yielded a $1\,\text{GeV}$ uncertainty for the Higgs mass. In the following we use this result.

\subsubsection{Matching at $M_{GUT}$}

In this section we explain how to obtain $\lambda(M_{SS})$ from a set of universal soft terms at the GUT scale,these are specified in \eqref{softterms}.
The  procedure to calculate $\lambda(M_{SS})$ goes as follows.
The values of the gauge and top Yukawa couplings in the NMSSM, $\hat{g}_1,\hat{g}_2,\hat{g}_3,\hat{y}_{t}$, are fixed by the corresponding $g_1,g_2,g_3,y_t$ in the SM via the matching conditions at $M_{SS}$.\footnote{The corresponding values of $\hat{y}_{t},\hat{g}_1,\hat{g}_2,\hat{g}_3$ used correspond to $m_h=125\text{GeV}$. Variations of $m_h$ between $50$ and $150\text{GeV}$ yield variations of $10^{-6}$ in the gauge couplings and $10^{-3}$ in the top Yukawa.} The matching conditions are given at tree level by
\be \hat{g}_1= \sqrt{\frac{5}{3}}g^{\prime},\quad \hat{g}_2=g_2,\quad \hat{g_3}=g_3,\quad\hat{y}_t=\frac{y_t}{\sin \beta}\,
\label{matchinggauge}\ee 
and receive one loop threshold corrections which can be found in \cite{Bagnaschi:2014rsa}. 
However, gauge couplings only enter in the calculation of the Higgs mass via the RGEs for the soft terms and thus, higher order corrections can be neglected.\footnote{ These could become important in precise estimations of gauge coupling unification.}
With $\hat{y}_{t},\hat{g}_1,\hat{g}_2,\hat{g}_3$ at hand we run from $M_{SS}$ up to $M_{GUT}$, with the RGEs of the NMSSM at two loops, to calculate $\hat{y}_{t_0},\hat{g}_{1_0},\hat{g}_{2_0},\hat{g}_{3_0}$. 
Using the latter boundary conditions for gauge and top Yukawa couplings together with \eqref{softterms} we implement the RGEs of the NMSSM again at the two loop level for Yukawa and gauge couplings and at one loop level for the soft parameters, to obtain the corresponding soft terms at $M_{SS}$. The RGEs are given in \cite{Ellwanger:2009dp} and we neglect the contribution of all Yukawa couplings except $\hat{y}_t$.\footnote{Neglecting the bottom Yukawa is a good approximation for low (or moderate) values of $\tan\beta$.} 
The soft parameters at $M_{SS}$ determine the values of $\tan\beta$ and $\delta$ via the eqs. \eqref{tanbeta} and \eqref{delta} respectively which yield $\lambda(M_{SS})$ through \eqref{lam}.\footnote{Notice that the value of $\tan\beta$ already appears in the matching condition of the top Yukawa given in \eqref{matchinggauge}, the values should of course match.}
 The one loop corrections given in \eqref{lam_threshold}, using the stop mixing in \eqref{stopmixing} are also included. With the value of $\lambda(M_{SS})$ at hand, we proceed as in \ref{matchingMSS}.

\section{The Higgs mass as a function of $M_{SS}$}\label{HiggsvsMss}

In this section we study the Higgs mass as a function of $M_{SS}$,
varying the NMSSM parameters \eqref{freepar} that determine the value of $\lambda(M_{SS})$ in \eqref{lam}. 
In section \ref{UVdependence} we proceed as before but
using the values of $\tan\beta$ and $\delta$ in $\lambda(M_{SS})$  computed from universal patterns of soft terms at the GUT scale. 

\subsubsection{$\lambda>0$ at SUSY-breaking scale}

\begin{figure}[h!]
\centering\includegraphics[scale=0.65]{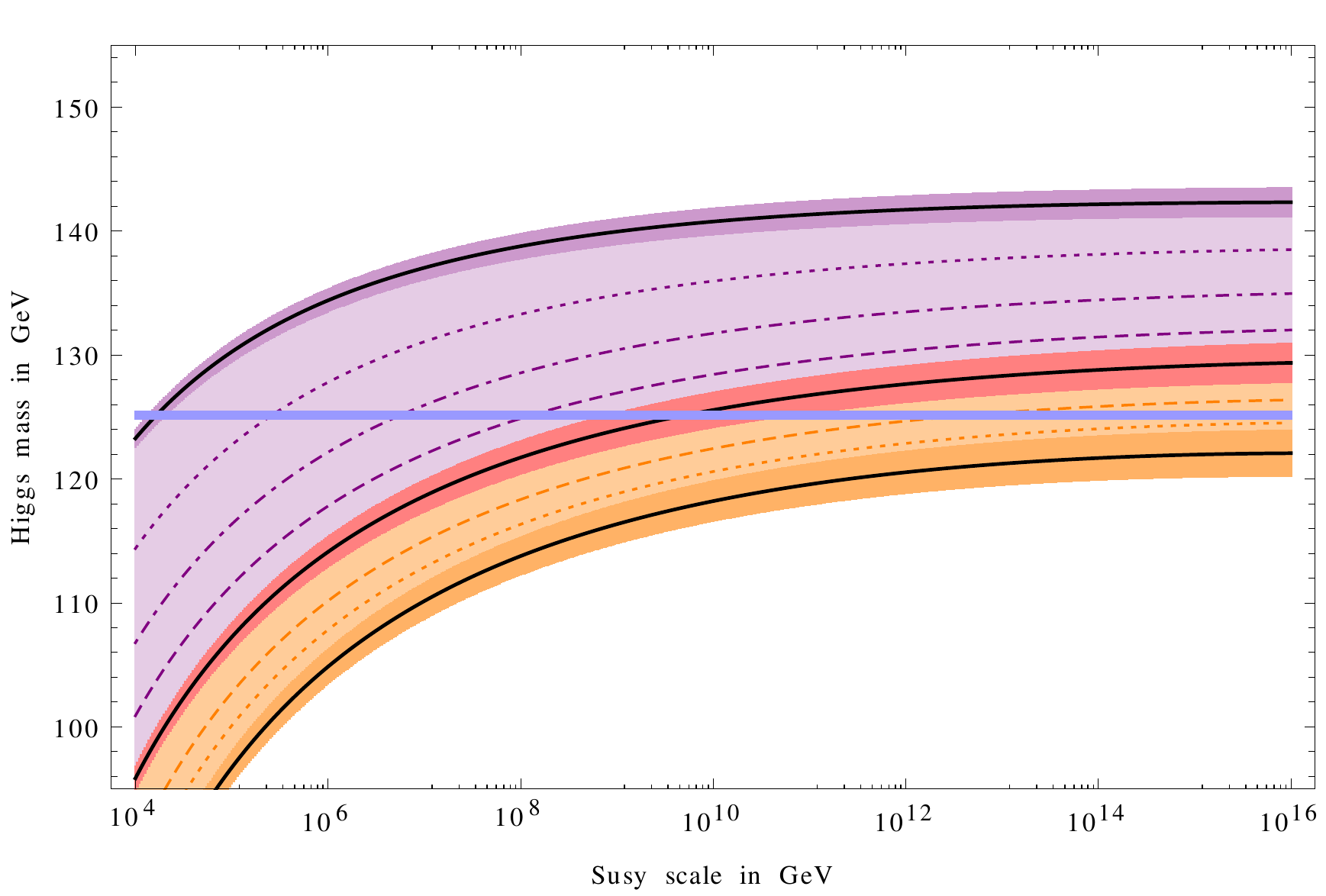}
\caption{Higgs mass as a function of $M_{SS}$ for $\tan\beta=1$. The region shaded in violet(orange) corresponds to $\lambda>0$($\lambda<0$) and from bottom-up $y_s=0.2,0.3,0.4,0.5$ ($y_s=0.3,0.25,0.2$ with $\delta=-2$). In red $\lambda=0$, $y_s=0$. We assumed $\tilde{X}_t=0$ and degenerate superparticles at $M_{SS}$. The bands display the uncertainty from varying $m_t=173.34\pm0.76 \text{GeV}$, we did not include them in all the curves to avoid clutter. The line in blue is the measured Higgs mass $125.15\pm0.25\text{GeV}$.}
\label{plot1}
\end{figure}

\begin{figure}[h!]
\centering\includegraphics[scale=0.65]{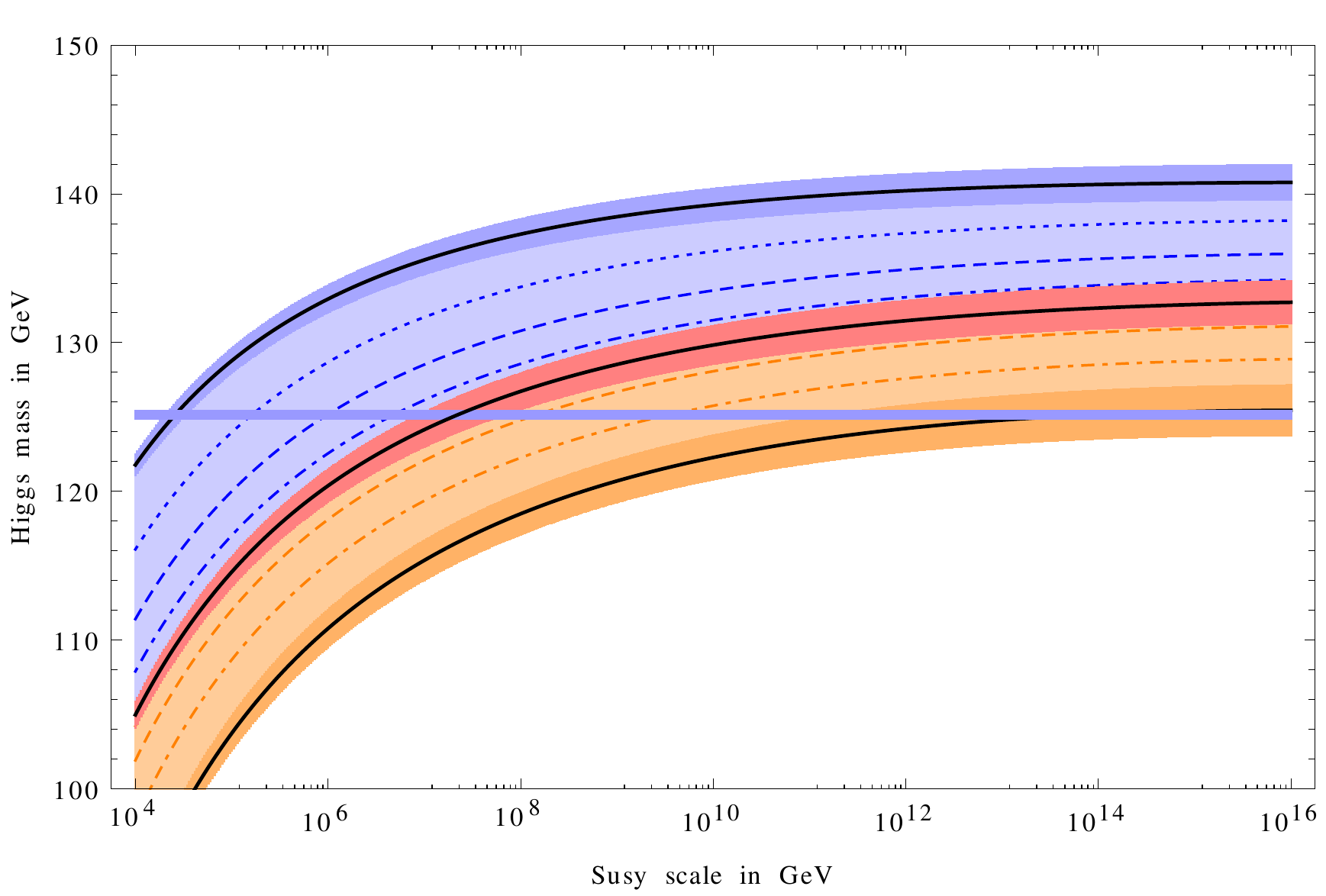}
\caption{Higgs mass as a function of $M_{SS}$ for $\tan\beta=2$. The region shaded in blue(orange) corresponds to, from bottom-up, $y_s=0.2,0.3,0.4,0.5$ ($y_s=0.4,0.3,0.2$ with $\delta=-2$). In red $y_s=0$. We assumed $\tilde{X}_t=0$ and degenerate superparticles at $M_{SS}$. The bands display the uncertainty from varying $m_t=173.34\pm0.76 \text{GeV}$. The line in blue is the measured Higgs mass $125.15\pm0.25\text{GeV}$.}
\label{plot2}
\end{figure}

To begin with, 
notice that 
the MSSM-like contribution to \eqref{lam} vanishes for $\tan\beta$ equal to one, while the NMSSM piece is maximized. 
From this observation we learn that
in the NMSSM low values of $\tan\beta$ can severely lessen the upper bound of $M_{SS}$ as $y_s$ increases.
In Figure \ref{plot1} and \ref{plot2} we plot the Higgs mass as a function of $M_{SS}$ for different values of $y_s$, assuming $\delta=0$ and $\tan\beta=1,2$ respectively. 
For large $\tan\beta$ the situation resembles the MSSM case \cite{Giudice:2011cg,Cabrera:2011bi}. In this regime the first term in \eqref{lam} is maximized and thus pulls $M_{SS}$ to low energies while the NMSSM term is almost vanishing regardless of the value of $y_s$.
 
\subsubsection{$\lambda<0$ at SUSY-breaking scale} 
 
As can be seen from \eqref{lam} by setting $\delta=0$, all universal contributions are strictly positive. This implies that independently of the values of the parameters in the supersymmetric theory, there exists a strict upper bound on $M_{SS}$ (within uncertainties) fixed by the scale $\mu_{0}$ at which $\lambda(\mu_0)$ vanishes, equivalently, the scale at which the SM becomes unstable. 
The state-of-the-art computation yields $\mu_{0}\simeq10^{10\pm1}\text{GeV}$ \cite{Buttazzo:2013uya}.
However, if $\lambda(M_{SS})$ can be matched to the expected negative value determined by the running of $\lambda$ in the SM, the
 upper bound on $M_{SS}$ disappears.\footnote{This statement assumes the metastability is not spoiled by the UV completion. A discussion upon this point can be found in \cite{DiLuzio:2015iua}.}  
In Figures \ref{plot1} and \ref{plot2} we plot the Higgs mass for various $y_s$, using $\delta=-2$ and $\tan\beta=1,2$ respectively. In these examples it can be clearly seen that $M_{SS}$ could take values up to $M_{\text{GUT}}$.

\subsection{Universal soft terms at $M_{\text{GUT}}$}\label{UVdependence}

As already anticipated, 
in this section we study the Higgs mass as a function of $M_{SS}$
assuming an underlying structure of universal soft terms at the GUT scale.  
In the NMSSM these are specified by the following parameters
\be 
m_0\,, M_0\,,A_0\,,\mu_{h_0}\,,\mu_{s_0} \text{ and }b_{s_0} 
\label{softterms}\ee
where $M_0$ denote the gaugino masses and the remaining soft terms are as introduced in section \ref{settingNMSSM}.\footnote{Notice that $m_3$ is fixed by the fine-tuning condition \eqref{ft}.} 
 The subindex $0$ indicates that the values are defined at the GUT scale, with $M_{GUT}=10^{16}\text{GeV}$.
Examples of universal soft terms in string theory can be obtained e.g. in the {\it dilaton domination} scenario \cite{Louis:2014pia}.

Notice from \eqref{tanbeta} that $\tan\beta$ is equal to one at the GUT scale and as $m_{h_u}^2,m_{h_d}^2\text{ and }\mu_h$ run, $\tan\beta$ evolves accordingly. However, $\tan\beta$ remains close to one for large values of $M_{SS}$. In other words, the unification (or universality) of soft masses predicts small values of $\tan\beta$, as already pointed out in \cite{Ibanez:2013gf}. For low values of the SUSY-breaking scale, $m_{h_u}^2$ becomes smaller and eventually changes sign, at this point there is a sudden increase of $\tan\beta$. 

On the other hand, $m_s,b_s\text{ and }\mu_s$ appearing in $\delta$ do not run for small values of $y_s$ and stay equal to their boundary conditions at the GUT scale. Thus, the running of $\delta$ is induced by $\mu_h$ and $A_{\lambda}$. As long as there are no hierarchies among the couplings, $\delta$ stays constant at large $M_{SS}$ and increases for low values of the SUSY-breaking scale.

In sum, two effects determine the Higgs mass in this scheme. At large values of $M_{SS}$, $\tan\beta$ takes very small values and so enhances the NMSSM correction to $\lambda(M_{SS})$. In this regime, $\delta$ does not significantly vary and, thus, the Higgs mass dependence upon $M_{SS}$ is (almost) constant. 
Furthermore, by tuning the value of $y_s$ near $\mathcal{O}(10^{-2})$, the Higgs mass can be easily accommodated in the experimental bound.
For lower values of $M_{SS}$, $\tan\beta$ starts to increase and thus $\lambda(M_{SS})$ becomes insensitive to the NMSSM correction. The latter competes with the MSSM contribution which grows with $\tan\beta$. The sum of these two terms leads to a slow decrease of the Higgs mass with $M_{SS}$. 

\begin{figure}[h!]
\centering\includegraphics[scale=0.65]{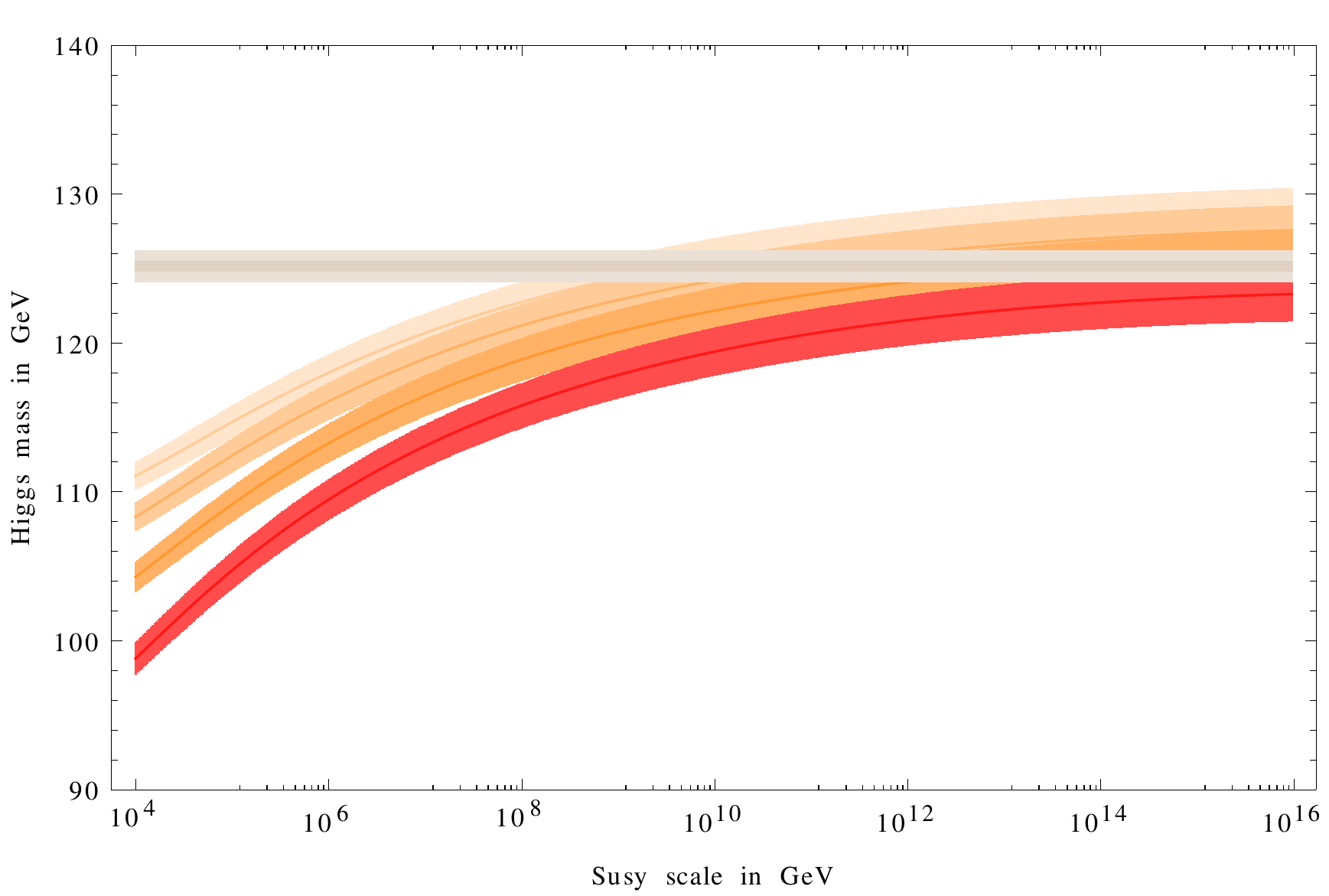}
\caption{Higgs mass as a function of $M_{SS}$ with soft terms at the GUT scale: $m_0=M_0=\mu_{s_0}$, $b_{s_0}^2=-\mu_{s_0}^2$ and $A_0=\mu_{h_0}=-1.5 M_0$  and $y_s=0.05\text{ (beige)},0.075\text{ (light orange)},0.1\text{ (dark orange)},0.125\text{ (red)}$. The bands correspond to the uncertainty bound in $m_t$ and the region in lighter brown is $m_h=125\pm1\text{GeV}$. In darker brown the experimental bound on $m_h$.}
\label{plot3}
\end{figure}

\begin{figure}[h!]
  \begin{minipage}[b]{.55\linewidth}
\includegraphics[scale=0.45]{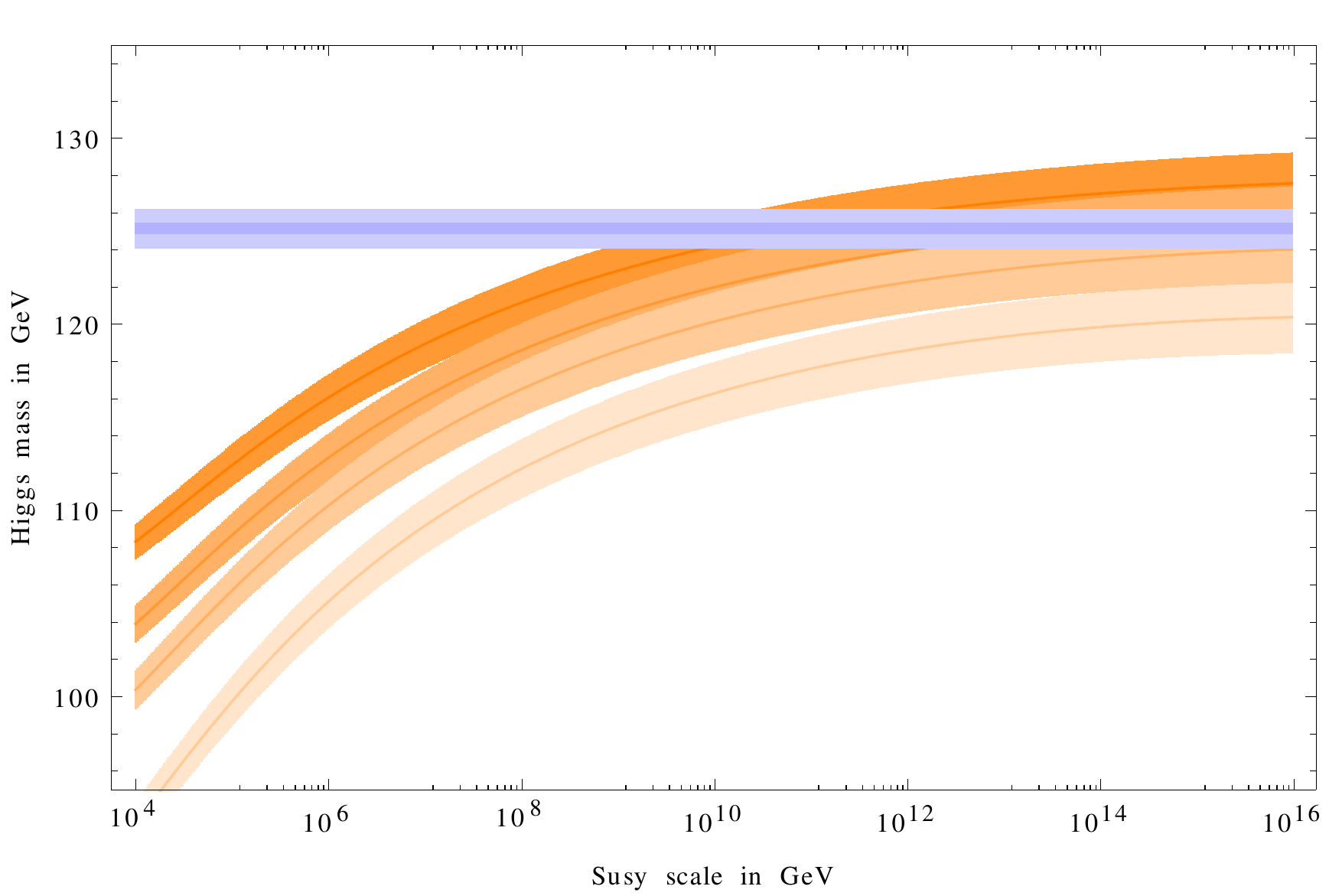}
  \end{minipage}%
  \begin{minipage}[b]{0.55\linewidth}
\includegraphics[scale=0.45]{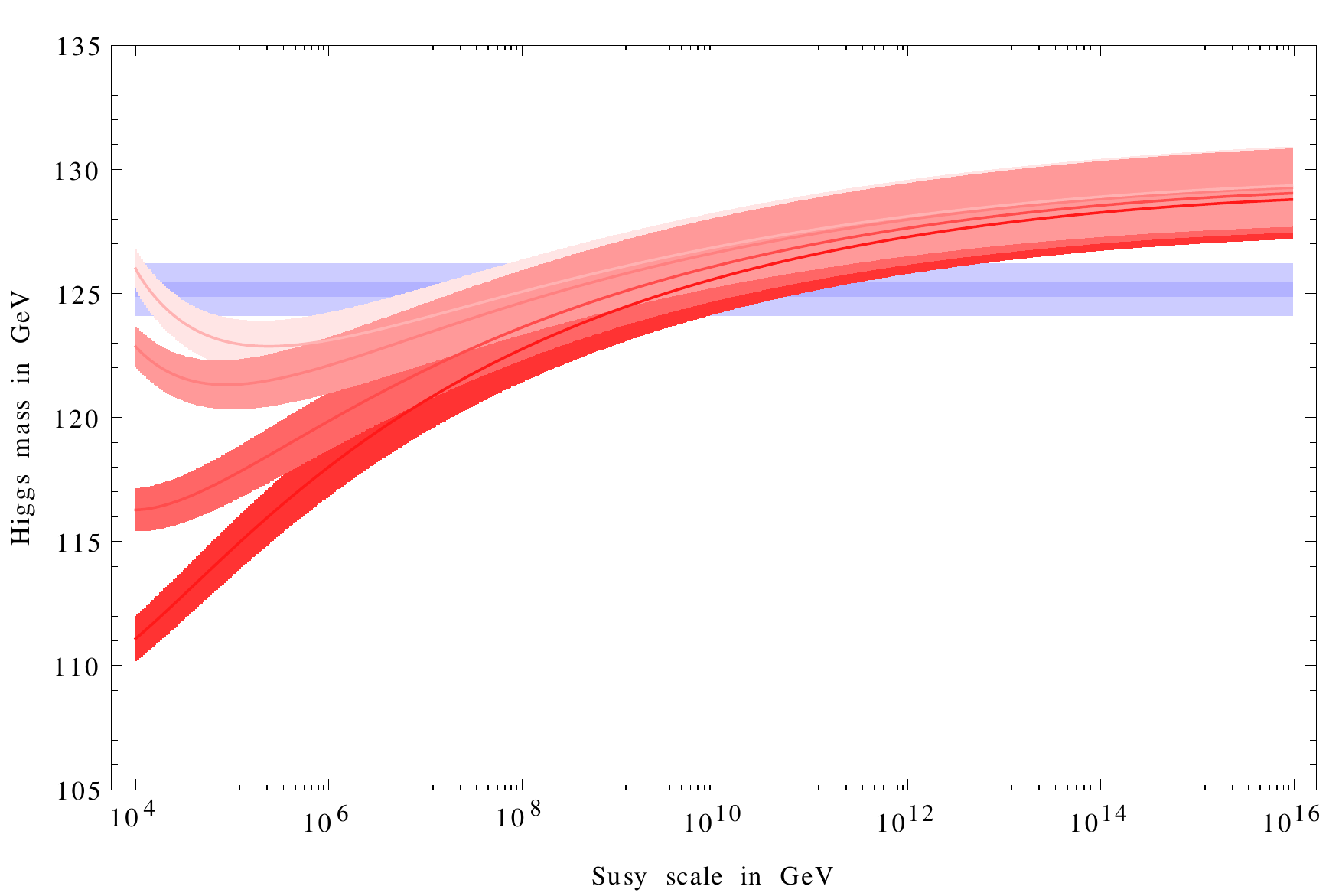}
  \end{minipage}
  \caption{ Higgs mass as a function of $M_{SS}$. The bands correspond to the uncertainty bound in $m_t$ and the region in blue is $m_h=125\pm1\text{GeV}$. In darker blue the experimental bound on $m_h$. ({\it left}) $y_s=0.075$ and (from dark to light orange) $m_0^2=M_0^2,\,0.6 M_0^2,\,0.45M_0^2 ,\, 0.3M_0^2$. ({\it right})  $y_s=0.05$ and (from dark to light red) $\mu_{h_0}=-1.5M_0,-1.25M_0, -M_0, -0.9 M_0$. 
}
\label{plot4}
\end{figure}

In Figure \ref{plot3} we show the Higgs mass as a function of $M_{SS}$ for the following choice of soft masses at the GUT scale, i.e. $m_0=M_0=\mu_{s_0}$, $b_{s_0}^2=-\mu_s^2$ and $A_0=\mu_{h_0}=-1.5 M_0$, and various $y_s$.\footnote{Notice that this choice of $A_0$ and $\mu_{h_0}$ minimizes the effect of stop mixing at large SUSY scales} 
As $y_s$ decreases the upper bound on $M_{SS}$ approaches the lower bound of $M_{SS}\simeq 10^{10}\text{GeV}$, that corresponds to the limit of $\tan\beta=1$ and $y_s=0$. 
Whereas increasing $y_s$ enhances the NMSSM negative contribution to $\lambda(M_{SS})$ and, thus,
pushes $M_{SS}$ to larger values. 
Similarly, in Figure \ref{plot4} (left), by fixing $y_s$ we show the effect of lowering $\lambda(M_{SS})$ by incrementing $\delta$. This can be achieved 
by taking smaller values for the soft masses.  


Notice that in the limit of vanishing $y_s$, the NMSSM contribution to $\lambda(M_{SS})$ is negligible. 
Hence, larger $\tan\beta$ values raise the Higgs mass at low energies. This effect is manifest depending on the choice of soft terms, in particular of $\mu_{h_0}$.\footnote{The choice of soft masses and A-terms, i.e. $m_0$ and $A_0$ have a milder effect on $\tan\beta$.}   
In Figure \ref{plot4} (right) we show the Higgs mass as a function of $M_{SS}$ for fixed $y_s$ and the same soft terms as before but varying $\mu_{h_0}$. 
As seen in Figure \ref{plot4}, for lower $\mu_{h_0}$, $\tan\beta$ becomes larger and boosts the Higgs mass.


\section{Conclusions}\label{conclusion}

In this letter we studied the Higgs mass in the NMSSM 
within High-Scale Supersymmetry.
In this setup, 
the low energy effective theory is described by the Standard Model and the respective couplings, in particular the quartic coupling $\lambda$, are fixed by the supersymmetric theory at the SUSY-breaking scale $M_{SS}$.
In the MSSM this matching sets $\lambda(M_{SS})$ to a positive (or vanishing) value and, thus, sets an upper bound $M_{SS}\lesssim 10^{10}\text{GeV}$. 
In the NMSSM, $\lambda(M_{SS})$ receives additional tree level corrections  given in \eqref{lam}.
The latter allow $\lambda(M_{SS})$ to take negative values,
 and, thus, the SUSY-breaking scale can take any value up to the GUT scale.
In this work, we computed the Higgs mass as a function of the SUSY-breaking scale varying the NMSSM parameters.
The results are summarized in Figures \ref{plot1} and \ref{plot2}.
For large values of $M_{SS}$, the Higgs mass becomes almost constant and can be easily adjusted to the experimental bound $125.15\pm0.25\text{GeV}$.

In addition, we studied special scenarios of supersymmetry breaking.
In particular, we assumed 
universal soft terms at the GUT scale to constrain the NMSSM parameters appearing in $\lambda(M_{SS})$. The unification of masses
yields low values of $\tan\beta$ at large $M_{SS}$ and, thus, enhances the NMSSM negative contribution to $\lambda(M_{SS})$. For
\be 10^{9}\text{GeV}\lesssim M_{SS}\lesssim 10^{16}\text{GeV}\,,\ee
the Higgs mass stays almost constant, and its value can be easily accomodated within the experimental bound.\footnote{
See section \ref{UVdependence} for details.} 
For lower values of $M_{SS}$ consistency with the Higgs mass becomes more model dependent and requires small NMSSM contributions to $\lambda(M_{SS})$.\footnote{In particular, $\tan\beta$ should take larger values and $y_s$ should be negligible in order to suppress the singlet contribution to $\lambda(M_{SS})$.} 
Furthermore, in this regime the soft masses spread out and, thus, one loop contributions to $\lambda(M_{SS})$ become large. Hence, for low values of $M_{SS}$ the determination of the Higgs mass relies on the details of the soft parameters.

To conclude, even if not directly accessible at the LHC, scenarios with large SUSY-breaking scales can be probed via the measured value of the Higgs mass. Furthermore, improved measurements of the top mass will help constraining the parameters, and in particular $M_{SS}$. 
Furthermore, as we have studied in this letter, the NMSSM can support SUSY-breaking scales near the GUT scale. 
These news are of particular interest for string phenomenology, since generic vacua in the landscape allow for supersymmetry breaking only at high scales.
It might be interesting to find explicit examples of the NMSSM from string derived setups. Provided specific patterns of soft terms, these models could be tested.

\section*{Acknowledgements}
I would specially like to thank Emanuele Bagnaschi, David Ciupke, Florian Staub, Alexander Voigt and Georg Weiglein for useful discussions. I would also like to thank Wilfried Buchm\"{u}ller, Jan Louis, Stefano di Vita and Alexander Westphal for interesting comments on the letter. This work is supported by the German Science Foundation (DFG) within the Collaborative Research (CRC) 676  "Particles, Strings and the Early Universe".

\bibliographystyle{hieeetr}
\bibliography{Bibliography}

\end{document}